\RequirePackage{fix-cm}

\newcommand{\stokes}{{\it s}}
\newcommand{\chan}  {{\it c}}
\newcommand{\base}  {{\it b}}
\newcommand{\rec}   {{\it r}}
\newcommand{\ns}    {${\rm{n}_{\rm s}}$}
\newcommand{\nb}    {${\rm{n}_{\rm b}}$}
\newcommand{\nc}    {${\rm{n}_{\rm c}}$}
\newcommand{\nr}    {${\rm{n}_{\rm r}}$}
\newcommand{\flagcal}{{\tt flagcal} }
\newcommand{\complex}{{\tt complex3}}
\newcommand{\tgmad}  {{\tt thrs\_gmad }}
\newcommand{\tpmad}  {{\tt thrs\_pmad }}
\newcommand{\tqmad}  {{\tt thrs\_qmad}}
\newcommand{\tngood}  {{\tt thrs\_ngood}}
\newcommand{\tblock}  {{\tt thrs\_block}}
\newcommand{\tvsr}    {{\tt thrs\_vsr }}
\newcommand{\tant}    {{\tt thrs\_ant }}
\newcommand{\tbase}   {{\tt thrs\_base}}
\newcommand{\tchan}   {{\tt thrs\_chan}}
\newcommand{\tclip}   {{\tt thrs\_clip}}
\newcommand{\tpeakt}  {{\tt thrs\_peak\_t}}
\newcommand{\tpeakf}  {{\tt thrs\_peak\_f}}
\newcommand{\tpeakx}  {{\tt thrs\_peak\_X}}

\newcommand{\wmint}   {{\tt win\_min\_t}}
\newcommand{\wminf}   {{\tt win\_min\_f}}
\newcommand{\wminx}   {{\tt win\_min\_X}}
\newcommand{\dwint}   {{\tt dwin\_t}}
\newcommand{\dwinf}   {{\tt dwin\_f}}
\newcommand{\dwinx}   {{\tt dwin\_X}}
\newcommand{\dpt}     {{\tt dpt}}
\newcommand{\dpf}     {{\tt dpf}}
\newcommand{\dpx}     {{\tt dpX}}
\newcommand{\nriterf} {{\tt nriter\_f}}
\newcommand{\nritert} {{\tt nriter\_t}}
\newcommand{\nriterx} {{\tt nriter\_X}}
\newcommand{\nsol}    {{\tt nsol}}
\newcommand{\nflag}    {{\tt nflag}}
\newcommand{\niter}   {{\tt niter}}
\newcommand{\salpha}   {{\tt alpha}}
\newcommand{\eps}     {{\tt eps}}
\newcommand{\refant}  {{\tt refant}}

\documentclass[smallextended]{svjour3}       
\smartqed  
\usepackage{graphicx,amsmath,multirow,natbib}

\journalname{Exp Astron}
\begin{document}

\title{FLAGCAL:A flagging and calibration package for radio interferometric data
}
\subtitle{}


\author{Jayanti Prasad \and
 Jayaram Chengalur         
}


\institute{Jayanti Prasad \at
              IUCAA, Postbag 4, Ganeshkhind, Pune University Campus, Pune 411007, India \\
             \email{jayanti@iucaa.ernet.in}           
           \and
            Jayaram Chengalur \at
            NCRA-TIFR, Postbag3, Ganeshkhind, Pune University Campus, Pune 411007, India
            \email{chengalur@ncra.tifr.res.in} 
 }

\date{Received: date / Accepted: date}

\maketitle

\begin{abstract}

We describe a flagging and calibration pipeline intended for making
quick look images from GMRT data. The package identifies and flags
corrupted visibilities, computes calibration solutions and interpolates
these onto the target source. These flagged calibrated visibilities
can be directly imaged using any standard imaging package. The pipeline
is written in ``C'' with the most compute intensive algorithms being
parallelized using OpenMP. 


\end{abstract}

\section{Introduction}
\label{sec:intro}

    Radio interferometric data taken at the GMRT has traditionally 
been analyzed interactively using the AIPS data package (but see
also \cite{sirothia09}). This becomes 
cumbersome when large data sets need to be analyzed, or when the data 
analysis has to be done in quasi real time. We describe here a "C" based 
program which calibrates GMRT data as well as flags data affected by 
interference or by instrumental problems. This package was developed
largely in the context of two ongoing programs at the GMRT viz. 
(a)~a search for transient radio sources for which quasi real time 
data analysis is needed and (b)~a search for HI emission at high 
redshifts for which large volumes of data have to be analyzed. The 
package is designed to give a quick look at the data for the first
program (in order to determine whether a given burst of radio 
emission can be localized in the sky or not) and to do the first 
pass of flagging and calibration for the second program. However,
since it operates on files in a standard format (viz. FITS)
and is relatively flexible, it is expected to be useful for a
larger range of problems than it was specifically designed
for.

    Radio interferometers measure ``visibilities'', i.e. the Fourier 
transform of the sky brightness distribution. Descriptions of the 
processes required to convert this information into an image of the 
sky can be found in \cite{thompson01,taylor99,chengalur03}. Here 
we focus only on the initial stages of this process, viz. that of 
identifying and flagging corrupted  data, and subsequently determining 
and correcting for the  antenna based complex gains.

Data at the GMRT could be corrupted for a variety of reasons,
e.g. instrumental failure, radio frequency interference, ionospheric
scintillations etc. A wide range of interfering signals could
be present, including interference that is narrow band and persistent
(e.g. from local digital equipment), interference that is broad band
but bursty in time (e.g. from satellites, aircraft radar etc.)
low level broad band interference that is persistent in time (e.g. 
from power transmission lines). A diverse range of approaches
have been taken for identifying interfering signals in radio
astronomical data (see e.g. \cite{chengalur96,kanekar97, urvashi03, 
briggs05, kocz10,offringa10,paciga11}). Here we identify corrupted 
data by assuming that the true visibilities should be continuous in 
the time-frequency plane. We identify discontinuities using robust 
estimators of the underlying statistics of the visibilities. 
This approach is well suited to finding RFI that is localised in time 
or frequency, but not low level persistent RFI. We also identify
corrupted calibrator visibilities by requiring that the visibility
phase for calibrators be stable over time.
We note that if the calibrator is resolved at some baselines or if there 
is confusing structure in the field of view, the requirement that the phase 
be stable will not be satisfied. However, we find in practice that at 
the GMRT operating frequencies and the generally used calibrators,
the visibility phase is stable enough for this algorithm to work.
Often data corruption affects entire subsets of data. For example, 
there may be narrow line RFI in one channel, or
a few contiguous channels. Similarly, a particular baseline may have
corrupted data because of a correlator problem. We identify such
subsets of corrupted data by making a second pass through the data.
Channels, baselines or antennas for which more than a specified fraction
of the data has been identified as corrupted in the first pass are
identified and then completely flagged out. We note that general 
purpose packages like AIPS and CASA also have tools that flag data
based on the assumption that the true visibilities should be continuous 
in the time frequency plane. While the algorithms that are used there 
are similar in spirit to those used here, \flagcal\ consistently uses
robust estimators and the most compute intensive calculations are
parallelized. It is also worth noting that a wrong choice of parameters
in algorithms that are based on smoothness in the time-frequency plane 
could lead to over flagging of the data.

Calculation of the robust statistics as well as of the calibration 
solutions is computationally intensive. Fortunately both calculations can
be trivially parallelized over time or frequency. 
Hence, wherever possible, these calculations are computed in parallel
using OpenMP pragmas. This gives substantial speed ups when running
the pipeline on the now standard multi-core workstations.

Below we give some details on the different algorithms used in the
program as well as present sample results and images made using data
processed through the pipeline.

\section{Details of the Pipeline}
\label{sec:pipeline}
\begin{figure*}
\begin{center}
\includegraphics[width=4in]{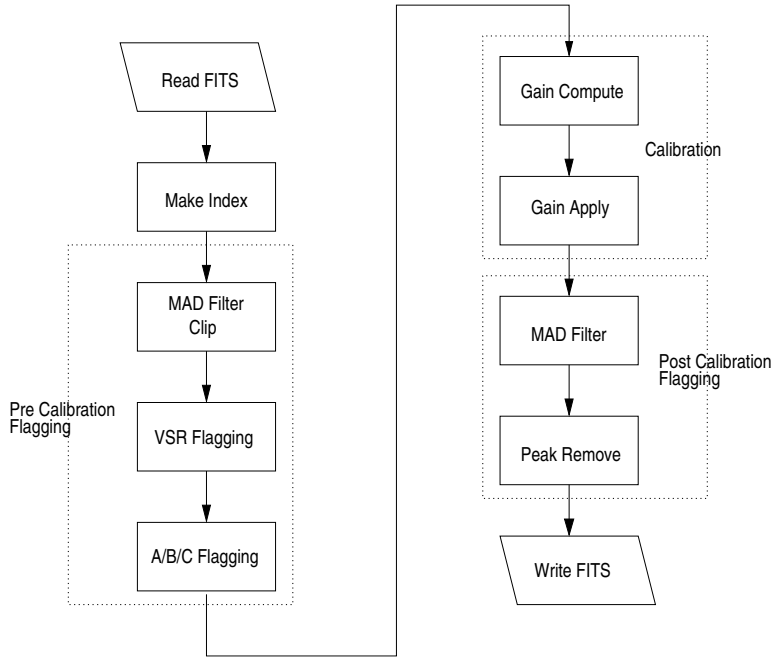}
\end{center}
\caption{ Flow chart for the \flagcal\ pipeline. See the text for more 
details.}
\label{fig:flowchart}
\end{figure*}

The flowchart for the \flagcal\ pipeline is shown in Fig.~\ref{fig:flowchart}.
The main stages of the pipeline are detailed below. The program consists of
a number of routines some of which take several user defined parameters 
(see Table~\ref{tab:pars}).
These parameters can be specified in a plain text configuration file
which is read by the program at run time. The configuration file also
allows the user to specify which of the routines are to be run on a given
data set, allowing a great deal of flexibility while executing
the program.

\subsection{Indexing and Pre Calibration Flagging}
\label{ssec:precal}

GMRT data is available as a random group UVFITS file. Accordingly,
the \flagcal program takes a random group UVFITS file as input and
its output is a calibrated and flagged random group UVFITS file.
The first stage of the pipeline reads the input random group fits file
and creates an index table for the file. The index table carries information
on the start and stop record, as well as the source ID and source type
(i.e. flux calibrator, phase calibrator or target source) for each
scan. By a scan we mean a continuous observation of a given source, with
fixed receiver and backend settings. A GMRT data file typically consists 
of a series of scans on different (e.g. flux calibrator, phase calibrator,
target source) sources. The visibility data themselves are stored internally 
in the form of a row-major multi dimensional array. The size of a typical 
GMRT data is  less than 4~GB. The current algorithm hence keeps
all of the indexed data in memory. However, for large files (greater than 8GB)
the algorithms can be easily modified to process the data scan by scan. This
would keep the memory requirements modest while also minimising the amount 
of disk I/O.
Each element of the array is a \complex (i.e. consisting of real, imaginary, 
and weight fields) structure.
Data with negative weight are regarded as being flagged. A typical visibility
data point for baseline \base, stokes parameter \stokes,\ frequency channel 
\chan\ and record \rec\ is indexed as  V[b +\nb(\stokes + \ns(\chan + 
\nc$\times$ \rec))], or $V_{\rec}(\chan,\stokes,\base)$ where \nb, \ns\  
and \nc\  are the total number of  baselines, stokes, and channels 
respectively.

  The next stage of the pipeline does pre-calibration flagging. Several
types of flagging are possible. The first is a simple thresholding or clipping,
where all visibilities whose amplitudes lie above a user defined 
threshold are clipped. If the user has some prior information on corrupted
baselines, antennas or channels, this can be passed to the pipeline in the
form of a flag file. Clipping and initial flagging 
(on the basis of the input flag file) is followed by two stages of 
MAD filtering\footnote{The median absolute deviation or MAD, which we refer
to as  $M_2$ in this paper, is defined by  the median of the absolute 
deviation around the median value.}. 
 A MAD filter  flags all visibilities whose amplitudes differ 
from the median amplitude by more than a user defined threshold times
the median absolute deviation (MAD). This filter has been chosen because 
of its robustness (i.e. the median and the median absolute deviation are 
robust to the presence of outliers, unlike the mean and the standard 
deviation). The two MAD filter steps before calibration are a Global~Mad~Filter 
and a Pre~Mad~filter. The Global~Mad Filter flags the data points 
for which the visibility amplitude is discrepant as compared to the 
the global median and MAD. These statistical parameters are computed over 
all baselines, channels and stokes parameters present in the data file for
that given source. For constructing the Global~Mad~Filter, the visibility 
data array $A_i$ where $i$ varies from  $0$ to 
\nr$\times$\nb$\times$\nc$\times$\ns\ is 
split into \nr\ sub-arrays, one for each time sample. The median 
$M_1^{j}$ and mad $M_2^{j}$ are computed for each sub array \footnote{
For an array $\{x\}=(x_1,x_2,x_3,.......)$ 
\begin{equation}
M_1 = \text{median}(x_1,x_2,x_3,....)
\nonumber 
\end{equation}
and
\begin{equation}
 M_2 = \text{median}(|x_1-M_1|,|x_2-M_1|,|x_3-M_1|,....)
\nonumber
\end{equation}
}.
For each source we collect the different $m_1^{j}$ and $m_2^{j}$ to 
construct the global median $M_1$ and MAD $M_2$ for that source.
All visibilities for source $k$ which satisfy
\begin{equation}
\text{abs}(|V_{\rec}(\chan,\stokes,\base)|- M_1^k) > \tgmad \times M_2^k
\label{eqn:gmad}
\end{equation}
are flagged, where  ~\tgmad~ is a user defined threshold 
(see Table~\ref{tab:pars}). Since the Global~Mad~Filter  takes into 
account all the data points available for a source, it is expected to 
be very robust.

In contrast to global mad filtering, pre~mad~filtering is done only 
on the basis of visibility data of the specific channel, stokes parameter
and scan. This filtering does not combine data from different channels, 
baselines or stokes parameters i.e., it works only in time domain.
In the Pre-Mad~Filter all visibilities that satisfy the following condition 
are flagged. 

\begin{equation}
\text{abs}(|V_{\rec}(\chan,\stokes,\base)|- M_1^k(\chan,\stokes,\base)) 
 > \tpmad \times M_2^k((\chan,\stokes,\base)
\label{eqn:pmad}
\end{equation}
where  \tpmad  is a user defined threshold value and 
where $M_1^k(\chan,\stokes,\base)$ and $M_2^k(\chan,\stokes,\base)$ are 
the median and mad values of the visibilities which are computed 
separately for every source $k$, channel \chan, baseline \base~ and 
stokes~\stokes.

\subsubsection{VSR and ABC flagging}
\label{ssec:abc}

For observations of an unresolved point source calibrator, one would expect 
that, in the absence of RFI or instrumental problems, the phase of the 
visibility for any given stokes, baseline and channel would vary slowly
with time. Consequently, for sufficiently small stretches of time, (typically
1-2 min at the GMRT) one would expect that the ratio of the amplitude of the 
vector sum of the visibilities to the scalar sum of the visibilities 
(the ``vector to scalar ratio'', VSR) would be close to unity.  In this 
stage of the pipeline, the calibrator visibilities 
are broken up into blocks (of size~~\nflag~~ decided by the user) and for 
each block the ratio ${\mathcal R}^j(\chan,\stokes,\base)$ define below
is computed.

\begin{equation}
{\mathcal R}^j(\chan,\stokes,\base) = \frac{ \left | \sum\limits_{r=r^j_{start}}^{t^j_{end}}
 {\vec V}_{\rec}(\chan,\stokes,\base)w_{\rec}(\chan,\stokes,\base) \right |}
{ \sum\limits_{r=r^j_{start}}^{r^j_{end}} |{\vec V}_{\rec}(\chan,\stokes,\base) w_{\rec}(\chan,\stokes,\base)| } 
\label{eqn:vsr}
\end{equation}

Where ${\vec V}_{\rec}(\chan,\stokes,\base)$ and $w_{\rec}(\chan,\stokes,\base)$
are the complex visibility and weight respectively for the flagging block $j$  
which starts at record $r^j_{start}$ and ends at record 
$r^j_{end}$. A flagging block is considered good only if it has a VSR
${\mathcal R}^j(c,s,b)$ that is greater than a user defined threshold ~\tvsr.
Clearly, the expected value of ${\mathcal R}^j(c,s,b)$ is unity for 
the case when the phase of the visibility remains constant and is
zero when the phase varies randomly between zero and $2\pi$.
We note that this test is applied only on the calibrator scans, and not 
on the target source scans. 

\begin{figure*}
\begin{center}
\includegraphics[width=4in]{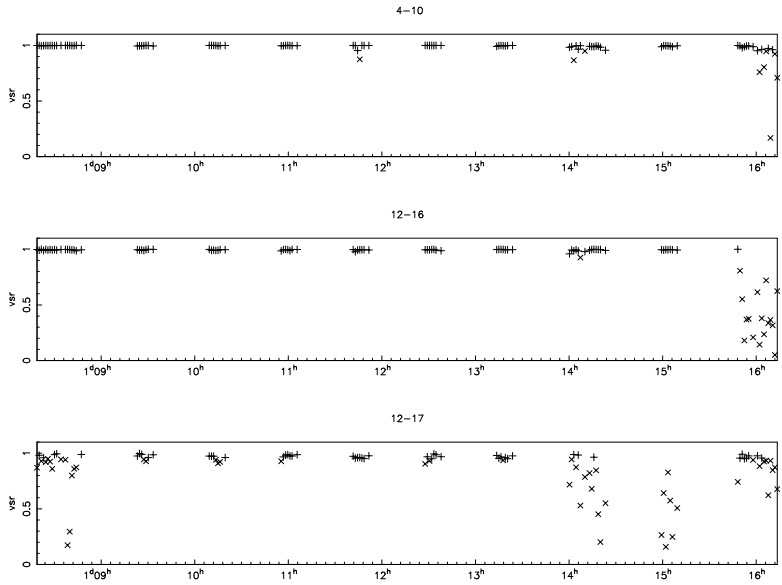}
\end{center}
\caption{Plots of the ratio of the modulus of the vector sum to the scalar sum
(VSR, see Eqn.~\ref{eqn:vsr}) of the visibility for the calibrator scans. For
good data the VSR is expected to be 1.0. All  data points for which 
the VSR is smaller than a user given threshold (0.95 in the above case) are
flagged and are shown by cross ($\times$) symbol. Data points which are
not flagged are shown by plus (+) symbol.} 
\label{fig:vsr}
\end{figure*}

In Fig.~\ref{fig:vsr} is shown the VSR for a typical frequency channels for  
few baselines. Data blocks for which the VSR is less than a user settable
threshold (0.95 in this case) are flagged. The default value of \nflag\ is 
2min. This value was iteratively determined after running the program on
several different GMRT data sets.

RFI or instrumental problems often affect a finite subset of the data, 
for example, a particular channel may be bad, or a particular baseline
or antenna. In order to identify such subsets of corrupted data, the VSR
${\mathcal R}^j(c,s,b)$  is marginalized over two of its indices to 
obtain a set of normalized  ``measures of the badness''
${\mathcal A, B}$ and ${\mathcal C}$ for all the antennas, baselines 
and channels respectively. Specifically, for each calibrator scan $i$ 
we compute:

\begin{align}
\mathcal{A}_l^i & = \frac{1}{N_A}
  \sum\limits_{c=1}^{n_{c}} \sum\limits_{s=1}^{n_{s}}   \sum\limits_{b=1}^{n_{b}}
{\mathcal F}^i(c,s,b) \left [\delta_D(p-l) + \delta_D(q-l) \right] 
\label{defa}
\\ 
\mathcal{B}_b^i &  = \frac{1}{N_B}
  \sum\limits_{c=1}^{n_{c}} \sum\limits_{s=1}^{n_{s}}  
{\mathcal F}^i(c,s,b)  \\ 
\mathcal{C}_c^i & = \frac{1}{N_C}
  \sum\limits_{b=1}^{n_{b}} \sum\limits_{s=1}^{n_{s}}  
{\mathcal F}^i(c,s,b) 
\end{align}

where the function $\delta_D$ is unity when its argument is zero, 
otherwise it is zero, $p,q$ are the indices of the antennas that comprise the baseline $b$.
We normalise ${\mathcal A, B}$ and ${\mathcal C} $ 
with $N_A = n_{c} \times n_{s} \times (n_{a}-1), N_B= n_{c} \times n_{s}$, 
and $N_C= n_{b} \times n_{s}$ respectively so that their values are
constrained to lie between zero and one, i.e. they represent the fraction
of data that has been identified as corrupted in the earlier pass 
through the data. ${\mathcal F}^i(c,s,b)$  takes values of either 0 or 1.
It is set to 0 only if one of the two conditions below are satisfied.

\begin{enumerate}
\item{The number of ``good blocks'' (i.e. blocks for which the value of 
${\mathcal R}$ is greater than the  user defined threshold ~\tvsr)  
is larger than a user defined threshold value ~\tngood. Note that~\tngood~
is in units of the flagging block length ~\nflag. Multiplying \tngood\ by
the time duration of the flagging block length would give the corresponding
time interval. }
\item{The length of the longest contiguous time interval in the scan 
 for which all the flagging blocks are ``good'' (i.e. as defined above)  
 is greater than a user defined threshold value \tblock.  As above, this
 is in units of the flagging block length.}
\end{enumerate} 

The idea behind the second test is that sometimes the problem causing the
data to be corrupted (e.g. hardware failure) occurs partway through the
scan. In that case there would be a contiguous stretch of time for which
the data quality is good. This can be distinguished from an intermittent
problem because if the problem is intermittent  the probability that 
nonetheless ~\tblock~ contiguous blocks remain unflagged is small. Note 
that we carry out the test~(2) on a scan only when the test~(1) fails. 
 
\begin{figure*}
\begin{center}
\includegraphics[width=4in]{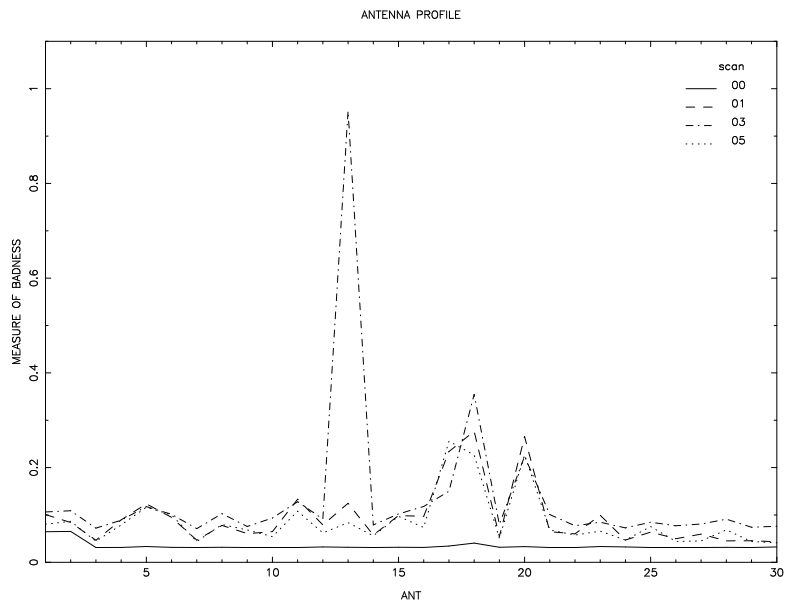}
\end{center}
\caption{\small\ This figure shows the measure of badness for antennas  
${\mathcal A}$  for different calibrator scans (shown by different line styles).
The value of  ${\mathcal A}$ is close to one for antennas which are 
"bad" and it is close to zero for antennas which are "good".}
\label{fig:badant}
\end{figure*}

\begin{figure*}
\begin{center}
\includegraphics[width=4in]{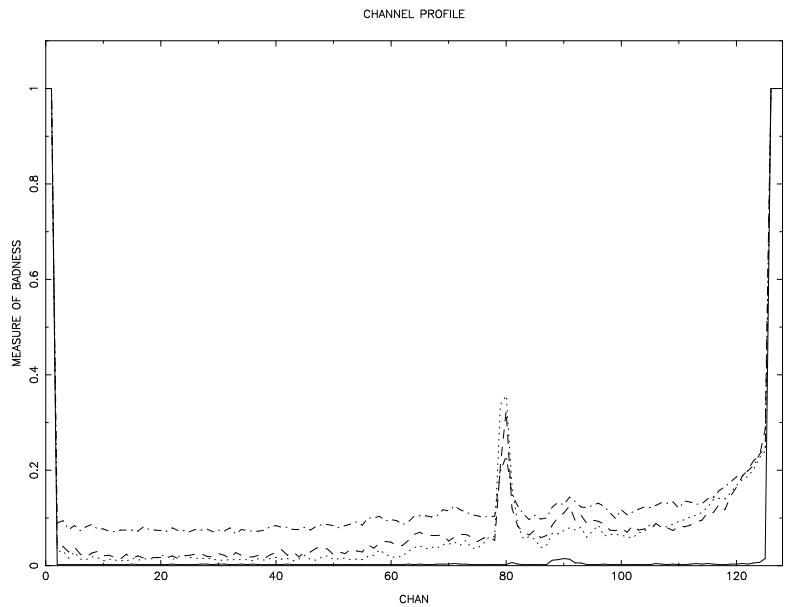}
\end{center}
\caption{\small\ This figure shows the measure of badness for different
channels ${\mathcal C}$  for different calibrator scans (shown by different 
line styles). The value of  ${\mathcal C}$ is close to one for channels
which are "bad" and it is close to zero for channels which are "good".}
\label{fig:badchan}
\end{figure*}

As mentioned above, we apply two passes of flagging for the calibrator scans, 
the first one based on ${\mathcal R}$ (where individual flagging blocks are
flagged) and the second based on  ${\mathcal A, B}$ and 
${\mathcal C}$ (where entire antennas, baselines or channels are flagged).
Antennas for which more than \tant\ of the data is already flagged, baselines
for which more than \tbase\ of the data is already flagged and channels
for which more than \tchan\ of the data is already flagged are flagged
in this second pass. Note that these thresholds represent fractions
of the total data for that antenna, baseline or channel.

For the target source, the assumption that the visibility
phase varies slowly with time need not be satisfied. However, it is
often the case that a data subset (e.g. a channel or a baseline) that
is corrupted during the calibrator scan, is also corrupted during the
target source scan. Hence the  user is allowed to interpolate the 
flags from the calibrator scans on to the target source scans. 
Specifically for the target source scans~$i$, the values of 
${\mathcal A, B,C}$ are computed using

\begin{equation}
X^i=max(X^{i-1},X^{i+1})
\end{equation}

where $X$ can be ${\mathcal A, B}$ or ${\mathcal C}$. The above interpolation 
scheme assumes that there is a calibrator scan before and after every target
source scan, which is typical for most GMRT observations. In case the target
source scan is not bracketed by calibrator scans, the interpolation is done
from the nearest calibrator scan. 

\subsection{Gain Computation and Calibration}
\label{ssec:gccal}

\subsubsection{Gain Computation}
\label{ssec:gain}

In situations where baseline based errors can be ignored, the relation between
the observed  visibility $V_{ij}$ and the true visibility $\widetilde{V}_{ij}$ 
can be written as: 

\begin{equation}
V_{ij} = g_i g_j^{*} \widetilde{V}_{ij}
\label{eqn:one}
\end{equation}

where $g_i$ is complex {\it gain}  (i.e. the overall gain, including 
instrumental and ionospheric contributions) of antenna $i$. The goal of a 
calibration 
program is to use the observed visibilities $V_{ij}$ to determine the gains
$g_i$ (which are expected to vary with frequency, stokes parameter and time)
and hence the true visibilities $\widetilde{V}_{ij}$. In ``ordinary
calibration'', which is the scheme implemented here, it is assumed that 
the {$g_i$}, , do not vary much with time. This allows one to compute
{$g_i$} using the observations of the calibrators (for which the true 
visibilities are known) and interpolate these on to the visibilities of
target source.

As is well known, Eqn.~\ref{eqn:one} describes an over constrained system 
since one has a total of $n_a(n_a-1)/2$ 
complex visibilities $V_{ij}$ from which one needs to determine $n_a$ 
complex gains. In the \flagcal\ implementation an iterative method of
solving for $g_i$ by least square minimization (see e.g.\cite {bhatnagar01})
is followed. The estimated complex gain of  $g_i^n$ at iteration $n$ is
given by:

\begin{equation}
g_i^n  = g_i^{n-1} 
+  \salpha \times  \left [ \frac{ \sum\limits_{j\ne i }^{n_a} X_{ij} g_j^{n-1} w_{ij}} 
{\sum\limits_{j\ne i}  |g_j^{n-1}|^2w_{ij} } -g_i^{n-1} \right ]
\label{eqn:calib1}
\end{equation}
where $X_{ij} = {V_{ij}}/{\widetilde{V}_{i,j}}$. We use the following convergence 
criteria for the above iterative solutions 

\begin{equation}
Max \biggl(\frac{ \left | g_i^{n} - g_i^{n-1} \right |}{|g_i^{n-1} |}\biggr) < \eps
\label{eqn:calib2}
\end{equation}

where the maximum is computed over all antennas. As formulated above the 
solutions $g_i$ are independent for each frequency
channel, stokes and record, and this algorithm can hence be trivially 
parallelized over any of them. In the \flagcal implementation the calculation
is parallelized over frequency channels. The solution is computed
after averaging the data over time; the length of the solution interval
~\nsol~and the maximum number of iterations ~\niter~ can be set by the user.
The typical solution interval is 2m and convergence is typically obtained
in less than 10 iterations. The computed gains compare well with those
computed independently by tasks in the AIPS package. The calibration
is estimated to be accurate at about the 10\% level. The residual 
errors are due to a combination of systematic errors (for e.g. any 
elevation dependent gains are not corrected for) and do not reflect 
the precision of the calculations.

\subsection{Post Calibration Flagging} 
\label{ssec:pmad}

After calibration, the visibilities on the calibrators should have identical
amplitudes and zero phase, within the noise. Data which is slightly corrupted
will hence be easier to detect than before calibration. We hence allow for
a stage of post calibration MAD filtering, in which all visibilities 
$V_{\rec}(\chan,\stokes,\base)$ for which

\begin{equation}
\text{abs}(|V_{\rec}(\chan,\stokes,\base)|-M_1(\chan,\stokes,\rec)) >  
\tqmad \times M_2(\chan,\stokes,\rec) 
\label{eqn:qmad}
\end{equation}

are flagged. \tqmad~~is a user defined threshold, and 
$M_1(\chan,\stokes,\rec)$ and $M_2(\chan,\stokes,\rec)$  are the median and mad 
for record $\rec$,  channel $\chan$, stokes $\stokes$ and baseline $\base$. 
The intention is that the calibration step can then be repeated, leading
to more robust gain solutions, although this is currently not yet implemented
in \flagcal. 

\subsubsection{SmoothSubtractThreshold (SST) Flagging}
\label{ssec:sst}

As discussed in Sec.~\ref{ssec:abc} the expected visibility for the target
source is not known a priori, and hence the flagging algorithms
suggested there may not be appropriate for target source visibilities.
However, even though one does not know the expected visibilities
for the target source, it is reasonable to assume that the visibilities
do not vary very rapidly with either time or frequency. One could then
try and identify corrupted data by either looking for deviant points after
polynomial fitting to the visibilities, or smoothing the data (see
e.g. \cite{chengalur96,kanekar97,urvashi03,offringa10}). In  \flagcal\ 
the latter approach is followed.
The data is smoothed over one dimension (either time or frequency) and then
visibilities that differ from the smoothed value by more than a 
threshold amount are flagged. This can be repeated for \nriterx\
iterations (where X can be time t or frequency f) with the smoothing
size increasing with each iteration. Specifically, for a set of $n$
visibilities ${x_i}$ one first computes the smoothed visibilities over 
a window of length $2w_j+1$ , viz.

\begin{equation}
{X}^j_i =
\begin{cases}
          & \frac{1}{2w_j+1}\sum\limits_{j=0}^{2w_j+1}x_{i+j}~~~~~~\text{for}~~0<~~i~~< w_j \\ 
          & \frac{1}{2w_j+1}\sum\limits_{j=i-w_j}^{i+w_j}x_{i+j}~~~~~\text{for}~~w_j~~<~~i~~<n-w_j \\ 
          & \frac{1}{2w_j+1}\sum\limits_{j=-(2w_j+1)}^{0}x_{i+j}~~~~\text{for}~~n-w_j~~<~~i~~<n 
\end{cases}
\end{equation}

then subtracts the smoothed data from the original 
\begin{equation}
{Y}^j_i=x_i - {X}^j_i 
\end{equation}

Data points for which
\begin{equation}
\text{abs}({Y}^j_i-M_1) > {\tpeakx}^j \times  M_2 
\end{equation}
where $M_1$ and $M_2$ are the median and median of absolute deviation for
the series ${X_i}$ are flagged. The user can choose a range of window
sizes $w_j$, and thresholds $T^j$ and the number of iterations $nriter\_X$ 
(see table~\ref{tab:pars} with $X$ replaced by $t$ for time and $f$ 
for frequency),  
over which this process can be repeated. As before this operation 
is parallelized using OpenMP pragmas. 

The width of the smoothing window $w^j$ and threshold ~$\tpeakx^j$\ 
at iteration $j$ is given by :

\begin{equation}
w^j=  (\dwinx)^j \times \wminx 
\label{eqn:peak1}
\end{equation}
and 
\begin{equation}
\tpeakx^j   = \left(\frac{1}{\dpx}\right)^j  \tpeakx  
\label{eqn:peak2}
\end{equation}
where descriptions of ~\wminx,~\dwinx,~\dpx~and ~\tpeakx~are 
given in Table~\ref{tab:pars} (with $X$ replaced by $t$ or $f$) with their
typical values. 

We note that running this procedure on the target source could result
in genuine emission from very strong transients being filtered out along
with undesired RFI. In such situations, it would be better to disable
this flagging. In the more general case, where one is searching for transients
that are too faint to show up in the visibility of a single channel of
a single baseline, this algorithm could be used.

\section{Results}
\label{sec:results}
\begin{figure*}
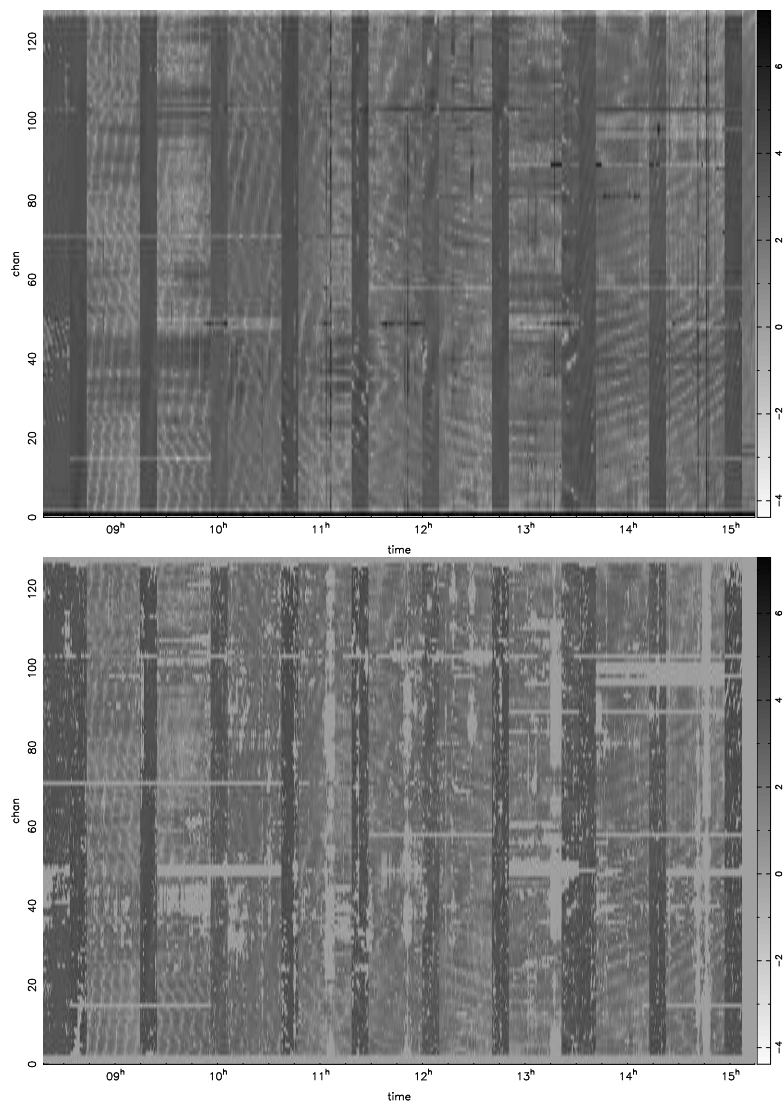

\begin{center}
\includegraphics[width=4in]{plot_gray1.ps} \\
\includegraphics[width=4in]{plot_gray2.ps}
\end{center}
\caption{[A] The top panel shows shows a time-frequency gray plot for a 
baseline for a typical GMRT observation at 157 MHz. The vertical stripes 
at constant intervals in the figure are the scans for phase calibrator. 
The presence of a large number of structures shows that the baseline is 
a heavily corrupted due to RFI. [B]~The bottom panel shows the same
data after processing through \flagcal. Note that the RFI has been
flagged out, and the intensity scale also has become narrower. The wedge
on right side shows the intensity scale in Jy.}
\label{fig:timefreq}
\end{figure*}

\begin{table} 
\begin{center}
\begin{tabular}{|l|c|c|c|c|} \hline \hline 
 Module name              &      1  &  2      & 4       & 8   \\ \hline  \hline
 Global Mad filter     & 205.68  & 108.08  & 66.93   & 41.45   \\ \hline
 Pre-Mad    filter     & 48.49   & 25.66   & 15.23   & 9.17   \\ \hline
 Gain computation      & 81.56   & 53.58   & 31.51   & 23.45   \\ \hline
 Peak in time          & 218.94  & 118.76  & 59.06   & 28.64   \\ \hline
 Peak in frequency     & 202.61  & 101.94  & 55.39   & 27.89   \\ \hline
 Post Mad filter       & 45.79   & 27.84   & 20.01   & 16.88   \\ \hline
 Total time            & 803.07  & 435.86  & 248.15  & 147.48  \\  \hline
\end{tabular}
\caption{This table shows the time taken in seconds by various modules
  for 1, 2, 4 and 8 threads (column 2, 3, 4 and 5 respectively). The
  last three rows show the total time taken (without considering the time take
  by read/write  and other serial sections). For timing we used a visibility 
  data set with 128 frequency channels, 2 stokes, 435 baselines and 1506 
  time samples  (16 second sampling). The file size is $\sim 2$GB and it
  takes $\sim 73$s to read in the file from disk. }
\label{tab:perf}
\end{center}
\end{table}

\begin{figure}
\begin{center}
\includegraphics[width=4in]{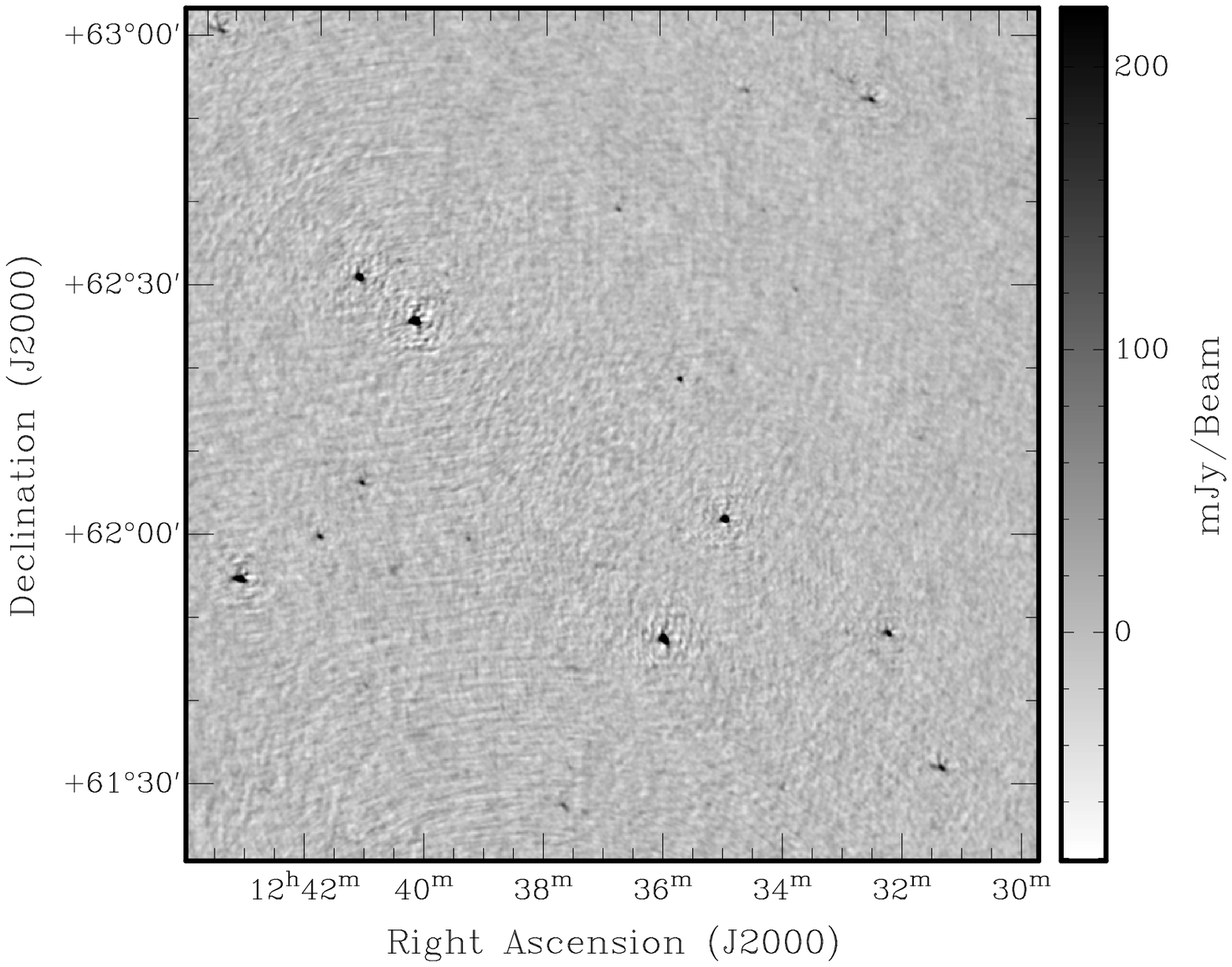}
\end{center}
\caption{A 157~MHz image obtained after processing the data with  \flagcal.
A snapshot of the processed visibilities from which this image has been
made is shown in Fig.~\ref{fig:timefreq}. Note that the only calibration
applied is that computed by \flagcal. The image was produced using the AIPS
task IMAGR and has been CLEANed, but not self calibrated. }
\label{fig:image}
\end{figure}

     The \flagcal\ programs have been tested several  GMRT data sets.
The program works well at removing RFI and calibrating for all the GMRT
bands with minimal tuning of user input parameters. The default values
of the parameters have been set after running the programs on various
data sets and examining what data was flagged. The fact that many of 
the parameters are expressed in terms of the underlying robust 
statistics of the data also reduces the necessity to fine tune them 
for each data set. In Fig.~\ref{fig:timefreq}
is shown a time frequency plot of visibilities on a given baseline
taken from a GMRT 150~MHz observation. This data set has been selected
for illustration because at GMRT the 150~MHz band is generally worst
affected by interference.  Panel~[a] shows the raw visibilities,
while Panel~[b] shows the visibilities after being processed through
\flagcal. As can be seen, the strong RFI visible in Panel[a] has been removed
and the calibration of the data also reduces the variation in the intensity.
About 30\% of the data has been flagged out, note that this also includes
data from antennas that had problems with their servo systems etc. 

An image made with this processed data is shown in Fig.~\ref{fig:image}.
The image was made using tasks in the AIPS package. No further calibration
or flagging was done in AIPS. Instead the data was imaged and cleaned
using the task IMAGR. The UVRANGE was set to 0-10 kL and the UVTAPER to
8~kL resulting in a resolution of $30^{''} \times 27^{''}$. Given that the 
pipeline is intended only for a quick look at the data we do not do more 
detailed cross checks. We postpone these for a later stage as the pipeline 
becomes more mature.

Table~\ref{tab:pars} lists the parameters available to the user in
tuning the performance of the \flagcal\ program. The execution times
for the various modules on an  AMD Opteron (2.6GHz, 8 GB RAM) with dual 
processors, each with 4 cores is given in Table~\ref{tab:perf}.

In summary, we present the initial results from \flagcal\, a pipeline
aimed at doing initial flagging and calibration of GMRT data. The aim
of this version of the pipeline was to pre process the visibility
data sufficiently in order to allow the user to make a quick look image
of the data. We demonstrate this using a GMRT 150~MHz data set, a frequency
band where RFI problems are generally severe. 

{\bf Acknowledgements} We are grateful to the referees of this paper
for their careful reading of the draft, and for their several useful
comments, which have substantially improved the readability of this
paper. This work was partially supported the Australia-India 
Strategic Research Fund.

\begin{center}
\begin{table*}
{\small
\hfill{}
\begin{tabular}{|l|l|l|}
\hline
\multicolumn{3}{|c|}{\bf\ Modules and their parameters} \\
\hline
Module name  & Parameter & Description of parameters (with reference)  \\ \hline
\multirow{5}{*}{Gain solution}   & \nsol (4)  & number of record used for gain solution (see \textsection\ref{ssec:gain})\\
 & \niter  (50) & number of iterations for gain solution (see \textsection\ref{ssec:gain})\\
 & \salpha (0.1) & a coefficients (see Eqn~(\ref{eqn:calib1})) \\
 & \eps  (0.01)  & used for checking  convergence (see Eqn~(\ref{eqn:calib2}))    \\ 
 & \refant  (1)  & reference antenna (which has phase zero)   \\ \hline 
\multirow{1}{*}{Clip } &  \tclip   (1000.0)& threshold used  for clipping   \\ \hline 
\multirow{1}{*}{Global mad filter } &\tgmad (9) & threshold used for filtering (see Eqn~(\ref{eqn:gmad}))  \\ \hline 
\multirow{1}{*}{Pre  mad filter } & \tpmad  (9)& threshold used for filtering (see Eqn~(\ref{eqn:pmad}))  \\ \hline 
\multirow{7}{*}{VSR/ABC flagging } &\nflag  (6) & number of samples used for VSR (see~\textsection\ref{ssec:abc} para 1) \\
                                  & \tvsr   (0.95)& threshold for VSR  (see Eqn~(\ref{ssec:abc}))   \\
                                  & \tngood (0.7) & threshold for bad block (see point~1~in~\textsection\ref{ssec:abc})  \\
                                  & \tblock (0.5) & threshold for bad block (see point~2~ in~\textsection\ref{ssec:abc}) \\
                                  & \tant   (0.3)& threshold for bad antennas (see Eqn~(\ref{ssec:abc}))    \\
                                  & \tbase  (0.3)& threshold for bad baseline  (see Eqn~(\ref{ssec:abc}))   \\
                                  & \tchan  (0.3)& threshold for bad channels  (see Eqn~(\ref{ssec:abc}))   \\ \hline 
\multirow{1}{*}{Post mad filter}  & \tqmad  (9)& threshold for post mad filtering (see Eqn~(\ref{eqn:qmad})) \\ \hline 
\multirow{5}{*}{SST flagging in time}& \nritert(4) & number of SST iterations (see  \textsection\ref{ssec:sst})  \\
                                  & \tpeakt (9.0)& starting threshold for SST (see Eqn~(\ref{eqn:peak2}))  \\
                                  & \dpt    (0.7)& factor for decreasing SST threshold (see Eqn~(\ref{eqn:peak2}))   \\
                                  & \wmint  (2)  & starting smoothing window for SST (see Eqn~(\ref{eqn:peak1}))  \\
                                  & \dwint  (2)  & factor for increasing SST window (see Eqn~(\ref{eqn:peak1})) \\ \hline 
\multirow{5}{*}{SST flagging in frequency}& \nriterf (4) & number of SST iterations  (see  \textsection\ref{ssec:sst})  \\
                                  & \tpeakf (9.0)& starting threshold for SST  (see Eqn~(\ref{eqn:peak2}))  \\
                                  & \dpf    (0.7)& factor for decreasing SST threshold  (see Eqn~(\ref{eqn:peak2}))   \\
                                  & \wminf  (2)  & starting smoothing window for SST  (see Eqn~(\ref{eqn:peak1}))  \\
                                  & \dwinf  (2)  & factor for increasing SST window  (see Eqn~(\ref{eqn:peak1})) \\ \hline 

\hline
\end{tabular}
}
\hfill{}
\caption{The Flagging and Calibration pipeline FLAGCAL consists of modules  which are used for
calibration and flagging. Every module has its set of parameters and the values of parameters 
control the effectiveness and accuracy of the module. In most cases the default (typical) values of the 
parameters can be used, however, in some case the tuning of parameters may be required. In the 
above table the fist column gives the names of modules, the second and the third columns list 
the parameters and give their description respectively. We also give the typical values of the 
parameters in the second column (in bracket) and the reference in the text where these parameters are 
discussed. The names of the parameters in the above table are identical to that are given in the code. Note that
the typical length of a GMRT records is 16 seconds.}
\label{tab:pars}
\end{table*}
\end{center}

\bibliographystyle{apj}
\bibliography{flagcal}
\end{document}